%% file: cas-dc-ASA-GS.tex
\pdfoutput=1 
\documentclass[a4paper,fleqn]{cas-dc}

\usepackage[authoryear,longnamesfirst]{natbib}
\usepackage{float}
\usepackage[linesnumbered,ruled]{algorithm2e}
\usepackage{caption}
\usepackage[section]{placeins}
\def\tsc#1{\csdef{#1}{\textsc{\lowercase{#1}}\xspace}}
\tsc{WGM}
\tsc{QE}
\begin{document}
\let\WriteBookmarks\relax
\def\floatpagepagefraction{1}
\def\textpagefraction{.001}

\captionsetup[figure]{labelfont={bf},name={Fig.},labelsep=period}
% Short title
\shorttitle{ASA-GS for Solving the CBPP-CI}

% Short author
\shortauthors{Y. Yuan, K. Tole, F. Ni, K. He et al.}  

% Main title of the paper
%\title [mode = title]{<main title>}  
\title [mode = title]{Adaptive Simulated Annealing with Greedy Search for the Circle Bin Packing Problem} 

% Title footnote mark
% eg: \tnotemark[1]
%\tnotemark[<tnote number>] 

% Title footnote 1.
% eg: \tnotetext[1]{Title footnote text}
%\tnotetext[<tnote number>]{<tnote text>} 

%% Group authors per affiliation:
\author[1]{Yong Yuan} 
\address[1]{School of Computer Science, Huazhong University of Science and Technology, Wuhan 430074, China.}
\fntext[co-authors]{The first two authors contribute equally.}
\fnmark[1]

\author[1,2]{Kevin Tole}
\address[2]{Institute of Computing and Informatics, Technical University of Mombasa, Mombasa 90420 - 80100, Kenya.}
\fnmark[1]

\author[1]{Fei Ni}

\author[1]{Kun He}
\cortext[correspondingauthor]{Corresponding author}
\ead{brooklet60@hust.edu.cn}
\cormark[1]

\author[1]{Zhengda Xiong}

\author[3]{Jinfa Liu}
\address[3]{Guangzhou Key Laboratory of Multilingual Intelligent Processing, Guangdong University of Foreign Studies, Guangzhou 510006, China}

\begin{abstract}
 We introduce a new bin packing problem, termed the circle bin packing problem with circular items (CBPP-CI). The problem involves packing all the circular items into multiple identical circle bins as compact as possible with the objective of minimizing the number of used bins. We first define the tangent occupying action (TOA) and propose a constructive greedy algorithm that sequentially packs the items into places tangent to the packed items or the bin boundaries. Moreover, to avoid falling into a local minimum trap and efficiently judge whether an optimal solution has been established, we continue to present the adaptive simulated annealing with greedy search (ASA-GS) algorithm that explores and exploits the search space efficiently. Specifically, we offer two novel local perturbation strategies to jump out of the local optimum and incorporate the greedy search to achieve faster convergence. The parameters of ASA-GS are adaptive according to the number of items so that they can be size-agnostic across the problem scale. We design two sets of new benchmark instances, and the empirical results show that ASA-GS completely outperforms the constructive greedy algorithm. Moreover, the packing density of ASA-GS on the top few dense bins is much higher than that of the state-of-the-art algorithm for the single circle packing problem, inferring the high quality of the packing solutions for CBPP-CI.
\end{abstract}

% Use if graphical abstract is present
%\begin{graphicalabstract}
%\includegraphics{}
%\end{graphicalabstract}

% Research highlights
\begin{highlights}
%\item We introduce a new bin packing problem termed the circle bin packing problem with circular items (CBPP-CI).
 \item First paper to introduce the circle bin packing problem with circular items (CBPP-CI).
\item Define a tangent occupying action and propose a greedy constructive algorithm for CBPP-CI.
\item Design two new operations, circle perturbation and sector perturbation, to generate neighbor solutions. 
\item Propose an adaptive simulated annealing algorithm with greedy search that obtains competitive results.
%\item Design two new operations that could escape from the local minimum trap and guide the search to reach a near-optimal solution.
\item Build two sets with a total of 52 new benchmark instances with 20 to 100 circular items. 
\end{highlights}

% Keywords
% Each keyword is seperated by \sep
\begin{keywords}
 \sep Packing \sep Heuristics \sep Tangent occupying action \sep Adaptive simulated annealing \sep Greedy search 
\end{keywords}

\maketitle

% Main text
\section{Introduction}
\label{sec:sec1}
\input{Sec1-Intro}

\section{Preliminary}%Problem Statement}
\label{sec:sec2}
\input{Sec2-Problem}

\section{Tangent Occupying Action Algorithm}
\label{sec:3}
\input{Sec3-TOA}

\section{Adaptive Simulated Annealing with Greedy Search}
\label{sec:sec4}
\input{Sec4-ASA-GS}

\section{Experiments}
\label{sec:sec5}
\input{Sec5-Exp}

\newpage
\section{Conclusion}
\label{sec:sec6}
In this paper, we introduce a new variant of bin packing problem termed the circle bin packing problem with circular items (CBPP-CI). For packing solutions, we define the tangent occupying action (TOA) to quickly pack the items into a bin as compactly as possible to minimize the number of bins used. Besides, we design a new form of optimization function embedding the number of bins used and the maximum density gap of the bins to evaluate the solution quality. 
We then propose the adaptive simulated annealing with greedy search (ASA-GS) algorithm to attain better solutions. The greedy search strategy can speed up the convergence rate. Based on the framework of simulated annealing algorithm, the parameters such as the times of greedy search, the acceptance probability are adaptive along with the number of items, which can help to sample the parameter space much more efficiently and attain a better solution for instances in a broad scale. To avoid getting trapped in local optimum, we propose two novel perturbation strategies, \textit{sector perturbation} and \textit{circle perturbation}. %The sector perturbation has its characteristic that the items are taken out and unassigned from the border to the bins' center. 
%The sector perturbation is more effective than the circle perturbation in most data sets.  
Experimental results show that ASA-GS exhibits good performance on the solution quality and computational time. Besides, the packing quality is better than that of the constructive algorithm TOA on all the CBPP-CI instances we generated. 
As this is a new problem, there is no baseline algorithm available. However, we see that the packing density of ASA-GS on the top few bins is much higher than the state-of-the-art results on the single circle packing problem, indicating the high quality of our solution.

%In the period ahead, we would like to extend our approach to address the three-dimensional sphere bin packing problem with spherical items (3D-SBPP-SI), which is a more challenging NP-Hard problem while has a great deal of real-life implementation.

% Numbered list
% Use the style of numbering in square brackets.
% If nothing is used, default style will be taken.
%\begin{enumerate}[a)]
%\item 
%\item 
%\item 
%\end{enumerate}  

% Unnumbered list
%\begin{itemize}
%\item 
%\item 
%\item 
%\end{itemize}  

% Description list
%\begin{description}
%\item[]
%\item[] 
%\item[] 
%\end{description}  

% Figure
% \begin{figure}[<options>]
% 	\centering
% 		\includegraphics[<options>]{}
% 	  \caption{}\label{fig1}
% \end{figure}

% \begin{table}[<options>]
% \caption{Problem Statement}\label{tbl1}
% \begin{tabular*}{\tblwidth}{@{}LL@{}}
% \toprule
%   &  \\ % Table header row
% \midrule
%  & \\
%  & \\
%  & \\
%  & \\
% \bottomrule
% \end{tabular*}
% \end{table}

% Uncomment and use as the case may be
%\begin{theorem} 
%\end{theorem}

% Uncomment and use as the case may be
%\begin{lemma} 
%\end{lemma}

%% The Appendices part is started with the command \appendix;
%% appendix sections are then done as normal sections
%% \appendix

%\section{}\label{}

% To print the credit authorship contribution details
%\printcredits

\section*{Acknowledgement}
This work is supported by National Natural Science Foundation (62076105) and Natural Science Foundation of Jiangsu Province (BK20181409).
%% Loading bibliography style file
%\bibliographystyle{model1-num-names}
\bibliographystyle{cas-model2-names}

% Loading bibliography database
\bibliography{mybibfile}

% % Biography
% \bio{}
% % Here goes the biography details.
% \endbio

% %\bio{pic1}
% % Here goes the biography details.
% \endbio
% \newpage
% ~~~~~~~~~~~~~~
% \appendix
% \setcounter{section}{0}\def\thesection{Appendix \Alph{section}}%
% \section{}
% \input{appendix-experiment}{}
\end{document}

%% file: Sec1-Intro.tex
As a classic combinatorial optimization problem, the packing problems aim to pack a certain number of items into one or multiple containers without overlapping. Most researches are for single container packing. The shape of the container can be rectangular, square, or circular, and the items can be rectangles or circles. As an important branch of operational research, the packing problems have a wide variety of applications in the logistic industry, circular cutting, container loading,
cylinder packing, etc. Meanwhile, it has been proved to be NP-hard by~\citep{demaine2010circle}. 
Hence there is no deterministic algorithm to find the exact solutions in polynomial time unless P = NP.

The bin packing problem (BPP) has been well studied for multiple container packing since the 1970s~\citep{johnson1973near}. There exist mainly two variants: the two-dimensional rectangular bin packing problem (2D-RBPP) and the two-dimensional square bin packing problem with circular items (SBPP-CI). The 2D-RBPP aims to pack a set of rectangular items into a minimum number of identical rectangular bins without overlapping~\citep{chung1982packing}. The impact of these techniques on the practical solution of 2D-RBPP has been quite impressive~\citep{christensen2017approximation}. For example,~\cite{KANG2003365} propose two greedy algorithms: IFFD and IBFD. IFFD assigns the items sequentially by the first-fit decreasing manner, and a new bin will be initialized when there is no more room for the packing; IBFD is a modification of IFFD, which assigns each item to the bin with the smallest remaining capacity. Other representative approaches include the tabu search~\citep{lodi1999heuristic}, the guided local search~\citep{faroe2003guided}, the hybrid GPASP/VND approach~\citep{parreno2010hybrid}, and various heuristics based on greedy method~\citep{lodi2002two, monaci2006set,wei2011skyline}. The SBPP-CI allocates all the circular items to a minimum number of square bins without overlap, which is first presented by~\cite{he2017greedy}. They further propose a greedy algorithm with corner occupying action to improve the packing quality by introducing the adaptive large neighborhood search~\citep{HE2021105140}.

To our knowledge, many studies have focused on multiple square or rectangular containers, while no significant published research addresses the problem of packing with multiple circular bins. Therefore, in this paper, we address a new variant termed the circle bin packing problem with circular items (CBPP-CI), which places a series of circular items inside multiple circular bins to minimize the number of bins used. It is an important extension of the two-dimensional circle packing problem (CPP), which is to pack all circular items into a single container of the circular or square shape to minimize the size of the container. Generally speaking, the approaches of CPP can be classified into two categories: constructive strategies and global optimization strategies. 

Constructive strategies sequentially pack the items into the bin based on some rules, such as the best-local position (BLP)~\citep{hifi2002best, hifi2004} and the maximal hole degree (MHD)~\citep{huang2006new}, which are defined to evaluate the benefit of a partial solution. Representative heuristics include the prune-enriched Rosenbluth method (PERM)~\citep{lu2008perm}, the augment beam search~\citep{akeb2009beam, akeb2011augmented}, the best-fit algorithm (BFA)~\citep{he2012efficient}, etc. %Moreover, some of them can be used to adjust the container size.

As the second category of approaches, global optimization strategies improve the solution iteratively based on the initial solution. It could be further subdivided into two categories: quasi-physical methods and meta-heuristic optimizations. The quasi-physical methods are based on a physical gradient or human-intuitive behavior to enhance the solutions obtained by problem-oriented heuristics~\citep{wang2002improved,lubachevsky1997curved}, while meta-heuristic optimizations usually have an evaluation function devised to employ a trade-off between randomization and local search, with the goal of directing and remodeling basic heuristics to generate feasible solutions. Typical algorithms include a simulated annealing approach
(SA)~\citep{hifi2004simulated}, monotonic basin hopping approach (MBH)~\citep{grosso2010solving}, iterated tabu search(ITS)~\citep{fu2013iterated}, action-space-based global optimization algorithm (ASGO)~\citep{he2015action}, formulation space search (FSS)~\citep{lopez2016formulation}, %the algorithm of partitioning narrow action spaces and circle items (PAS-PCI)~\citep{hePAS},
adaptive tabu search and variable neighborhood descent (ATS-VND)~\citep{ZENG2018196}, etc. 

Most of the constructive solutions focus on the traditional CPP and are designed on the specific characteristics of the problem. These methods are no longer applicable for CBPP-CI because of the characteristic gap between CPP and CBPP-CI. Moreover, although the global optimization technique can be used on CBPP-CI as a general search framework, it lacks adaptive adjustments, including the search strategy and evaluation function. Otherwise, the search efficiency is poor, and it is hard to find an iterative optimization method to make further improvements based on the current solution.

%Most of the constructive solutions are usually incomplete, and it is challenging to devise an evaluation function to judge the solution quality. On the contrary, the global optimization technique always generates complete solutions that facilitate weighing the solution quality by devising an evaluation function. However, this approach is quite hard to make further improvements based on the current solution. 

As the CBPP-CI is a new problem, there are no available benchmark instances. Following our previous works on the square bin packing problem with circular items (SBPP-CI) in~\citep{HE2021105140, he2017greedy}, we choose two categories of benchmarks for the single circle packing problem (SCPP) on the packomonia website \footnote{\url{www.packomonia.com}} and build two sets of new benchmark instances based on them for the CBPP-CI. For the solving method, we first propose a greedy heuristic based on the designed tangent occupying action (TOA), which can quickly obtain a competitive packing result. TOA always places the current circular item tangent to any two packed items or the bin boundary. At the same time, we also need the packing item to have a minimum distance to the bin boundary. In this way,  items are packed as compact as possible, and the remaining space can all gather in the center area of a bin. To judge whether an optimal solution has been found, we continue to design adaptive simulated annealing with greedy search (ASA-GS) method inspired by related works~\citep{HE2021105140,hifi2004simulated,geng2011solving}.
In contrast to the TOA algorithm, we apply a globalization approach that improves the packing pattern iteratively. We first present an energy function to be minimized and offer an initial packing solution. Then we try to seek more adaptive parameter control to improve the solution quality on large-scale instances. Besides, we utilize the greedy search strategy to achieve faster convergence. Finally, to avoid falling into local optimal solutions, we propose two novel perturbation strategies, and the experiments have verified their effectiveness. Moreover, the packing density of ASA-GS on the top few bins is much higher than the best results for the single circle packing problem on the packomonia website, which indicates the high quality of our solution.

The main contributions of this work are summarized as follows:
\begin{itemize}
\item We address a new and important variant of BPP termed CBPP-CI, which comprises packing circular items into multiple circle bins as compactly as possible to minimize the number of used bins. Moreover, we build two sets of new benchmark instances for CBPP-CI.%based on the single container circle packing problem for the CBPP-CI.
\item We propose a constructive greedy algorithm based on the devised tangent occupying action that can quickly generate a competitive solution.
\item We define an energy function for simulated annealing and present two novel perturbation methods (\textit{sector perturbation} and \textit{circle perturbation}) to generate neighbor solutions. Besides, we incorporate a greedy search to achieve faster convergence. %The experiments demonstrate their effectiveness.
\item The parameters are adaptive along with the number of items such that our algorithm can obtain the better solution for the CBPP-CI with a broad scale. 
\end{itemize}

The rest of this paper is organized as follows: Section~\ref{sec:sec2} presents a formal definition of the CBPP-CI and our alternate optimization function, which could help find denser packing so as to minimize the objective. Section~\ref{sec:3} gives some definitions and proposes the constructive algorithm. Section~\ref{sec:sec4} presents two perturbation operators and describes the ASA-GS algorithm in detail. Section~\ref{sec:sec5} shows and analyzes the experimental results. Section~\ref{sec:sec6} concludes the work with future work recommendations.

%% file: Sec2-Problem.tex
In the proposed circle bin packing problem with circular items (CBPP-CI), we are given $n$($n \in N^{+}$) circular items $C_{1}$, $C_2$, \dots, $C_n$ with radius $r_{1}$, $r_2$, \dots, $r_n$, and a set of $n$ identical circular bins with radius $R$ (w.l.g. for any circular item $C_{i}$, $r_i \leq R$), we aim to determine the center coordinates of each item $C_{i}$ in a bin such that all items are packed feasibly, i.e. with all circular items fitting completely inside the bins and no overlapping exists between any pair-wise items (i.e.($C_i \cap C_j = \varnothing$)). The goal is to minimize the number of used bins, denoted as $K$ $(1\le K\le n)$.

\subsection{Problem Formulation} 
\label{sec:2.1}
Assume that the center of each circular bin $B_{k}$ is located at $(R,R)$ in two-dimensional Cartesian coordinate system and denote the center of each circular item $C_i$ as $(x_i,y_i)$. We can define a packing solution as $X = \{<x_1, y_1, b_1>, <x_2, y_2, b_2>, \dots,  <x_n, y_n, b_n>\}$, where $b_i$ is the indicator that the placement of item $C_i$ in the $b_{i}$-th bin $B_{b_i}$ ($b_i \in \{1, \dots, K\}$). In order to formulate the problem, a summary of necessary variables is listed in Table \ref{tab:f1}. 

\begin{table}
\centering
\caption{Variable definition.}\label{tab:f1}
\begin{tabular*}{\tblwidth}{@{}LL@{}}
\toprule
  Variable & Description\\
\midrule
$n$ & Number of circular items\\
$C_{i}$ & The $i$-th circular item\\
$r_{i}$ & Radius of $C_{i}$\\
$\left(x_{i},y_{i}\right)$ & Center coordinates of $C_{i}$\\
$B_{k}$ & The $k$-th bin\\
$R$ & Radius of the circular bins\\
$I_{ik}$ & Indicator of whether $C_{i}$ is in the $k$-th bin\\
$Y_k$ & Indicator of whether the $k$-th bin is used \\
$d_{ij}$ & Distance between points $(x_{i},y_{i})$ and $(x_{j},y_{j})$\\
\bottomrule
\end{tabular*}
\end{table}

The CBPP-CI problem can be formalized as minimizing $K$ while satisfying the following constraints: 
 \begin{equation} 
    \sum_{k=1}^{n} I_{ik}=1,\label{eq:2.1}
 \end{equation}
where $I_{ik}\in \{0,1\}$ and $i,k\in\{1,\ldots,n\}$, implying that each circular item is packed exactly once.
CBPP-CI also requires that any pair-wise items in the same bin (i.e. $I_{ik}=I_{jk}=1$, $\forall i,j,k\in \{1,\ldots,n\}$) must not overlap:
\begin{equation} 
d_{ij}=\sqrt{(x_{i}-x_{j})^{2}+(y_{i}-y_{j})^{2}}\ge (r_{i}+r_{j})I_{ik}I_{jk}.\label{eq:2.2} 
\end{equation}
Third, to ensure that every circular item is placed entirely inside a bin, CBPP-CI requires:
\begin{equation}
\sqrt{(x_{i}-R)^{2}+(y_{i}-R)^{2}} + r_{i} \le R.
\label{eq:2.3}\end{equation}
Finally, we use $Y_{k}$ to indicate whether there exist circular items packed into a bin $B_{k}$: 
\begin{equation} 
Y_{k}=\left\{ \begin{array}{ll} 1,\,\textrm{if}\,\sum_{i=1}^{n}I_{ik}>0,\,i,k\in \{1,\ldots, n\},\\0,\,\textrm{otherwise}.\end{array}\right.
\label{eq:2.4} 
\end{equation} 
And the goal is to minimize the summation of $Y_{k}$: 
\begin{equation} 
\min K=\sum_{k=1}^{n}Y_k,\label{eq:2.5} 
\end{equation}
and clearly $1\le K\le n$.

We could associate the items in bin $B_k$ as an item set, denoted as $S_k$.
%, then the packing will group the circular item $C_i$ into set $S_{k}$ if $b_i$ is equal to $k$. 
So a solution can be obtained by two steps: we first partition the items into different sets $\mathcal{S} = \langle S_{1}, S_{2}, \dots, S_{K} \rangle$ for the bins; then we try to pack the items of $S_{k}$ into bin $B_k$ without overlapping. An optimal packing is that the number of bins used can not be reduced any further.

%\subsection{Objective Function} 
\subsection{Optimization Function} 
\label{sec:2.2}
The overall goal of the CBPP-CI is to use as few bins as possible to pack the $n$ circular items $C_{i}$, %, where $i\in\{1,\ldots n\}$, and it has been summarized 
as shown in Eq.~(\ref{eq:2.5}).
%
%\begin{equation} 
%\min K=\sum_{k=1}^{n}Y_k.\label{eq:2.5} 
%\end{equation}
%
However, to attain the global optimum, it is necessary to consider a more local objective function that focuses on packing as tightly as possible. In this regard, suppose that a packing solution $X$ corresponds to a partition $S=S_{1}\cup S_{2}\cup\ldots\cup S_{K}$ such that $S_{k}$ is the set of circular items that are packed in bin $B_{k}$, and $k\in\{1,\ldots, K\}$. Let $A$ be the area of a bin (all bins are identical). Then, the density of packing $S_{k}$ into a bin $B_{k}$ is given by:
\begin{equation} 
d_{B_{k}}(X)=\frac{1}{A}\sum_{C_{i}\in S_{k}}\pi  r_{i}^{2},  \qquad \text{where~~} A = \pi R^{2}.\label{eq:2.6}
\end{equation}

Given a packing solution $X$ and $k\in\{1,\dots,n\}$, let $d_{\min}=\min\{d_{B_{k}}(X)|1\le k \le K \}$ and $d_{\max}=\max \{d_{B_{k}}(X)|1\le k\le K\}$. A useful local optimization function is defined as follows:
\begin{equation} 
v(X)=d_{max}-d_{min}.\label{eq:2.7}
\end{equation}
The greater the value of $v(\cdot)$, the higher the quality of a feasible solution $X$. Since an increment in $v(\cdot)$ corresponds to a tighter packing as some items move from sparser bins to the denser bins.

Further, we need to minimize the value of $K$, i.e., to maximize the value of $-K$.  
%\begin{equation} 
%\text{min~~}K \equiv \text{max~~}(-K).\label{eq:2.8}
%\end{equation}
So we define our optimization function as:
\begin{equation} 
\text{max~~} F(X)=  -K +  d_{max}-d_{min}.\label{eq:2.9}
% F(X)=  -K +  d_{max}-d_{min}.\label{eq:2.9}
\end{equation}
The greater the value of $F(\cdot)$ is, the better and tighter the packing is. %Hence, our goal becomes:

Note that $0\le d_{max}-d_{min} \le1 $, this term is used for regularization. It implies that the optimization function is more inclined to use fewer bins, and the difference in the number of bins is enough to weigh different solutions. When two feasible packings use the same number of bins, we will focus on each candidate solution's densest bin and the sparsest bin. The denser the densest bin is, the less the wasted space is. The more sparse the sparsest bin is, the more concentrated and complete the remaining still-reserved space is, making it easier to pack the following circular items. Therefore, we assume such a difference in density could determine the quality of candidate solutions.

%In order to construct the problem for circular bins, we use a similar setup. As before, let the circles $C_{i}$ be ordered by their integer radii ${\it r}_{i}$ so that ${\it r_{1}},  {\it r_{2}},  \ldots, {\it r_{n}}$, where, generally, ${\it r_{i}}\ge {\it r_{i+1}}$. %

%Equations (\ref{eq:2.1})--(\ref{eq:2.3}) are the constraints for the CBPP-CI. As an equivalence to condition
%For equation (\ref{eq:2.3}), we will refer to a circular item $C_{i}$ %as being in a \emph{reference position}, if the distance from the %center of $C_{i}$ to point $(R,R)$\HK{check} is in the interval %$(0,R-r)$. 

%A feasible packing solution to the CBPP-CI is represented as $X$,  and through the packing $X$, we can see the circular item $C_i$ is assigned to which bin. 

%% file: Sec3-TOA.tex
This section introduces the concept of tangent occupying action and then proposes a constructive greedy algorithm based on this action. We want to pack circular items into the bins as compact as possible through the tangent occupying action to reduce the number of bins used.

\subsection{Definitions}
We first provide several essential definitions, especially the tangent occupying action.

\textbf{Definition 1. (Tangent occupying action)}.
A tangent occupying action (TOA) is a packing action that chooses an outside circular item to place to a position inside a bin such that the item is tangent to any two or more packed items (the circular bin can be regarded as a special hollow item).

%\newline\textbf{Definition 2. (Feasible packing position)}. A packing position of a circular item in a bin is feasible if it does not violate any constraints; i.e., items must not overlap and must be fully contained in a circular bin. (See Eq. (\ref{eq:2.1})--(\ref{eq:2.5}) for detailed constraints). 
%

\textbf{Definition 2. (Quality of a feasible packing position)}. For an item, the quality of a feasible packing position is determined by the distance between the center of the packing item and the circular bin's boundary: 
\begin{equation} 
d\left(x,y\right)= R - \sqrt{(x-R)^2 + (y - R)^2} - r,
\label{eq:3.1}
\end{equation}
where $(x,y)$ is the center of the circular item. The smaller the interval, the better the packing position.

All feasible positions are sorted in the ascending order of $d(x,y)$ for a circular item in the current bin. 
 %In lexicographical order, 
A smaller $d(x,y)$ is better, which allows more concentrated free space in favor of placing the remaining circular items. The idea is to pack circular items nearer to the bin's boundary.

%For a circular item in the current target bin, all feasible positions in the bin are arranged in the ascending order of $d(x,y)$. A smaller $d(x,y)$ means the packing position is closer to the edge of the bin. We recommend that the circle be placed closer to the edge of the bin, which will allow the free space to be more concentrated and facilitate the placement of subsequent circles.

\subsection{TOA Algorithm}

\begin{algorithm}[htbp]
\caption{TOA Algorithm}
\label{alg: TOA-Algorithm}
\SetKwInOut{Input}{Input}
\SetAlgoLined
\SetKw{KwBreak}{break}
\Input{A vector of unassigned circle's ID: circle\_ids, a vector of bin's ID: bin\_ids, bin's radius: $R$;}
\KwResult{For each circle $C_i$, find a bin $B_k$, and place the circle center at $\left(x_i,y_i\right)$;}
\For{$i \in circle\_ids$}{
    $vector <TOA> s = \varnothing$ \; 
    $bin\_id\_idx = 0$\;
    \While{true}{
        \If{$bin\_id\_idx == bin\_ids.size()$}{
            return false\;
        }
        $s\leftarrow$ Compute feasible packing positions for $C_i$\;
        \If{$s\neq \varnothing$}{$\KwBreak$\;}
        $bin\_id\_idx \leftarrow bin\_id\_idx + 1$; \tcp{Turn to the next bin}
    }
    TOA  $best\_toa =$ Select the best packing position from $s$ with $d(x,y)$\;
    $circles[i].x = best\_toa.p.x$\;
    $circles[i].y = best\_toa.p.y$\;
    %$bins[bin\_ids[ bin\_id\_idx]].insert(i)$\;
    Place the circles[i] into the bin\_ids[ bin\_id\_idx] bin;
}
\end{algorithm}
Details of the TOA algorithm are presented in Alg.~\ref{alg: TOA-Algorithm}. It works by packing circular items sequentially in a particular order of their radii (e.g., from large to small). To load the current item, we first locate all the TOAs of the first bin that satisfies the problem constraints. If there is no available TOA, we seek the next bin to continue searching feasible TOAs until at least one available TOA occurs. Among all possible TOAs, we select the placement with the minimal distance $d(x,y)$ and place the item at $(x,y)$ in the current bin. The TOA algorithm iterates the above procedure until all circular items have been loaded into the bins without overlapping. With this process, TOA prefers positions closer to the bin's boundary. Hence, it packs the circular items as compact as possible and utilizes the bin space greedily to minimize the number of bins used.

TOA is very fast in constructing a solution, but it could not obtain a solution with excellent quality. Therefore, we present two novel mutations and introduce a meta-heuristic global optimization approach called ASA-GS to improve the solution quality.

%% file: Sec4-ASA-GS.tex
 %In this section, We explore a hybrid meta-heuristics to tackle the problem of CBPP-CI. According to the characters of CBPP-CI, the Sector perturbation and Circle perturbation are proposed to help the stagnated packing pattern escape from the local optimum and explore other solution space. Further, we present an adaptive simulated annealing algorithm with the greedy search(ASA-GS) that strengthens the packing quality through perturbation on an initial packing solution. It can speed up the convergence rate and be suitable for the problem on a broad scale by making the parameters adaptive.
%
Simulated annealing (SA) algorithm ~\citep{kirkpatrick1983optimization} has been extensively developed and widely used in many optimization problems. It can avoid getting trapped in the local optimum and attain better solutions by accepting worse solutions with a certain probability. To strengthen the packing solution, we propose a boosted algorithm called the adaptive simulated annealing with greedy search (ASA-GS) for the CBPP-CI. Our method is inspired by the works~\citep{HE2021105140,hifi2004simulated,geng2011solving} that can guide the algorithm quickly converging to optimal solutions. 

The ASA-GS algorithm~\citep{geng2011solving} is described in Alg.~\ref{alg: ASA-GS}. In ASA-GS, there are several decisions to be made: how to define the energy function $f(\cdot)$; how to attain an initial solution; how to generate a neighbor solution; how to determine the assignments of parameters such as the probability of accepting a new solution, and the current temperature.

In what follows, we show how one can use the principle of the ASA-GS algorithm to solve the CBPP-CI.

\subsection{Energy Function}% associated with the CBPP-CI}
\label{sec:subsec4.1}
According to our defined optimization function of the packing problem, we define the energy function $f(\cdot)$ as $-F(\cdot)$ for the simulated annealing algorithm:
\begin{equation}
    f(X) = -F(X) = K - d_{max} + d_{min}.
    \label{eq:4.1}
\end{equation}
It can be seen from Eq.~(\ref{eq:4.1}) that minimizing the energy function $f(\cdot)$ is equivalent to maximizing the optimization function $F(\cdot)$. Therefore, the smaller the value of $f(\cdot)$, the better a packing solution.

%\begin{itemize}
%\item[1.] Initialize the annealing parameters;
%\item[2.] Determine the initial solution $X$;
%\item[3.] generate a new solution $X^{'}$;
%\item[4.] if $dE \le 0$ or $ e^{-dE/T_k} \ge random(0,1)$, then set $X = X^{'}$;
%\item[5.] update the value of parameters such as the coefficient of temperature,the probability of accept a new solution.
%\item[6.] repeat steps 3-5 until N times or satisfy exit criterion.
   
%\end{itemize}

\begin{algorithm}[htbp]
\caption{ASA-GS Algorithm}
\label{alg: ASA-GS} 
\SetKwInOut{Input}{Input}
\SetKwInOut{Output}{Output}
\SetAlgoLined
\SetKw{KwBreak}{break}
\Input{Bin radius $R$, a set of $n$ circular items $\{C_i | 1\leq i\leq n\}$ with radii $r_1, \ldots,r_n$~ $(r_i \geq r_{i+1})$}
\Output{A dense packing solution $\mathbf X$ for CBPP-CI. }
Initialize the annealing parameters $t_{start}$, $t_{cool}$, $N$, $t_{greedy}$, and set $t_{current} = t_{start}$, $G = 0$ \;
Initialize a packing solution $\mathbf {X_{0}}$ and let $\mathbf{X = X_{0}}$\;
\For{$i\leftarrow 1 $ \KwTo $N$}{
    Select one perturbation method between sector perturbation and circle perturbation\;
    Compute $dE = f(X^{'}) - f(X)$\;
    \tcp{See Subsection~\ref{sec: subsubsec4.3.3}  and Algorithm~\ref{alg: Generate New Solution} for details}
    $\mathbf{X}^{'} \leftarrow $ Generate a new packing solution$(X,R)$\;
    
    \eIf{$dE \leq 0$}{
       $X = X^{'}$;\tcp{Accept the new solution}
    }
    {
        $G = G + 1$ and compute $f(X^{'}_{G})$\;
        \eIf{$G \geq t_{greedy}$}{
        Select $X^{'}_{best}$ with condition $f(X^{'}_{best}) = min(f(X^{'}_{1}),f(X^{'}_{2}),...,f(X^{'}_{t_{greedy}}))$ \;
        %$X^{'} = X^{'}_{best}$\;
        \tcp{Accept the best solution with probability $p$}
        \If{$e^{(-dE/t_{current})\times log(n/2) } >= rand(0,1)$}{
            $ X = X^{'}_{best}$\;
        }
       
     }
     {
     Continue to generate next neighbor solution\;
     }
    }
    $t_{current} = t_{current} \times t_{cool}$ and let $G = 0$\;
    \If{$t_{current} \le t_{end}$}{break\;}
    }
\end{algorithm}

\subsection{Initial Packing Solution}
\label{sec:subse4.2}
\begin{figure*}[htbp]
\centering
    \includegraphics[width = 1\textwidth]{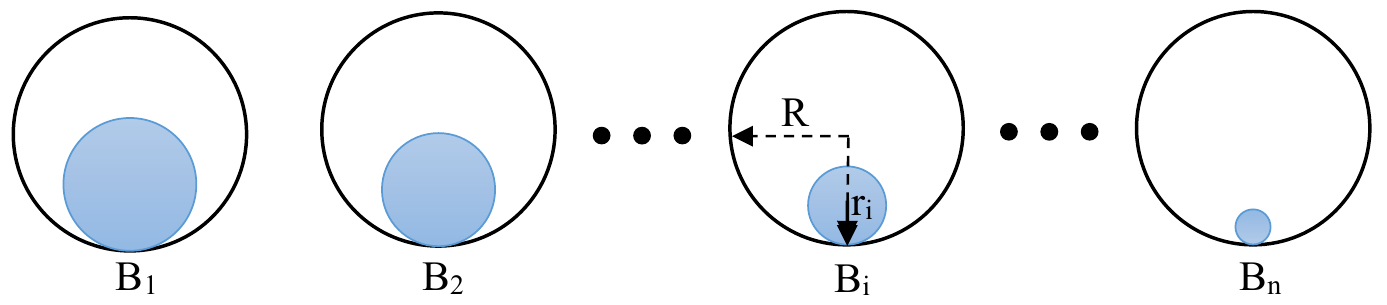}
    \caption{Initialize a packing solution.}
    \label{fig:initial packing}
\end{figure*}

We can easily obtain an initial packing solution using $n$ circular bins and assigning each circular item $C_{i}$ in bin $B_{i}$ as shown in Fig.~\ref{fig:initial packing}.
\begin{equation}
\begin{split}
\textbf{Initialize\_packing\_solution}\left(P\right)=\\\left\{\left\langle R,r_{i},B_{i}\right\rangle | i\in\{1,\ldots,n\} \right\}
\end{split}
\end{equation}

% The initial solution also can be quickly constructed by the greedy algorithm TOA (Alg.~\ref{alg: TOA-Algorithm}).
% \begin{equation} 
% \textbf{compute\_initial\_solution}(P)= GACOA(\{{C_{i}}|1\leq i \leq n \},R)
% \end{equation}

\subsection{Generate Neighbor Solutions} %with Different perturbation} %strategies}
\label{sec:sec4.2}
Generally, a new neighbor solution is obtained by conducting a local disturbance to the current solution.
%, which can substantially shorten the time. Therefore, 
Here an effective perturbation strategy plays a significant role in heuristic algorithms in the local search process to solve the optimization problem. Besides, different perturbation methods usually have different impacts on the specific problem. To void falling into local optimum, we design two new perturbation strategies for the CBPP-CI, termed circle perturbation and sector perturbation.
%  %\begin{comment}
% \begin{figure*}[pos=h]
% %\centering
% \begin{minipage}[t]{0.5\linewidth}	
% \includegraphics[width=0.9\textwidth]{fig/Circle perturbation.png}	
% \caption{Circle perturbation} 
% \label{fig: circle perturbation}
% \end{minipage}
% \hfill
% \begin{minipage}[t]{0.5\linewidth}
% \includegraphics[width=0.9\textwidth]{fig/Sector perturbation.png}
% \caption{Sector perturbation} 
% \label{fig: sector perturbation}
% \end{minipage}
% \end{figure*}
% %\end{comment}

\subsubsection{Circle perturbation}
As Alg.~\ref{alg: SampleCircles-algorithm} shows, the circle perturbation strategy selects a circular item randomly in a circular bin $B_k$, then generates a circular area with the item's center as its center, the radius of the circular area is a random number in $[0,\frac{R}{2}]$. It guarantees that at least one item will intersect the generated circular area. In most cases, more than one item will cross this area and be reassigned at each iteration.
\begin{figure}[htbp]
\centering
\includegraphics[width=0.35\textwidth]{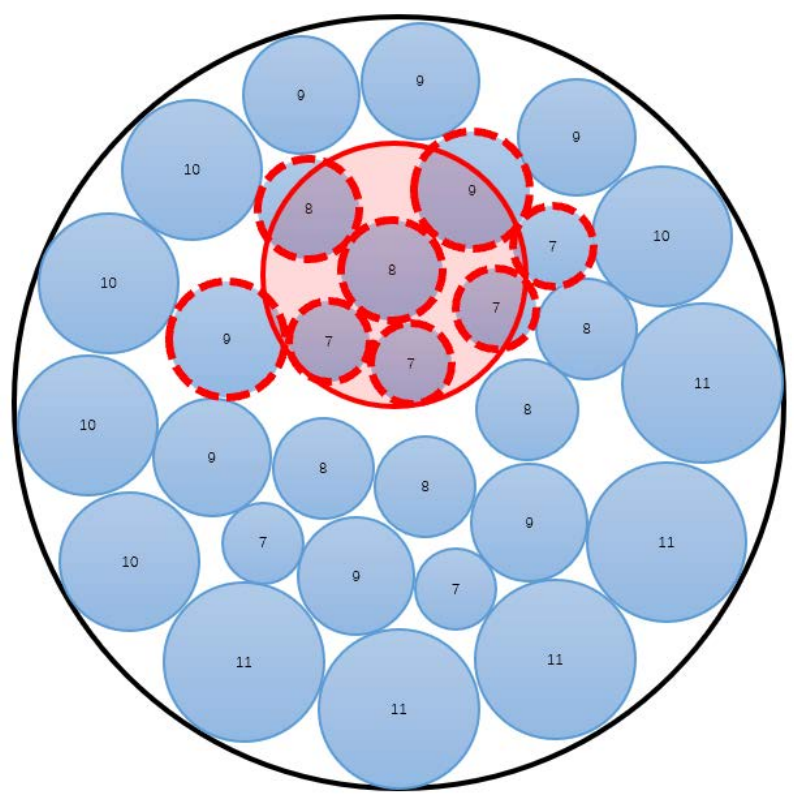}	
\centering
\caption{An illustration of circle perturbation.} 
\label{fig: circle perturbation}
\end{figure}

%%
%%%%%%%%%%%%%%%%%%%%%%%%%%%%%%%%%%%%%%%%%%%%%%%%%%%%%%%%%%%%%%
\begin{algorithm}[htbp]
 \caption{Pseudo-code of sampling a circle}
 \label{alg: SampleCircles-algorithm}
 \SetKwInOut{Input}{Input}
 \SetKwInOut{Output}{Output}
 \SetAlgoLined
 \SetKw{KwBreak}{break}
 \Input{Bin $B_{k}$, bin radius $R$}
 \Output{A circular area with $<x,y,r>$}
 \tcp{Each circle is represented as $<x,y,r>$}
 $r\leftarrow\mathrm{random\_real}(R / 2) $; \tcp{The circle radius is r}
 \tcp{Randomly select a circular item from $B_{k}$}
 \If{$ (!B_k.empty())$ }{
  $i \leftarrow\mathrm{random\_ints}(1,\left\{ i\left|C_i \in\mathbf{B_k}\right.\right\})$;}
  $x \leftarrow C_i.x$\;
  $y \leftarrow C_i.y$\;
  $circle = Circle(x,y,r)$;\tcp{Generate a circle area}
\end{algorithm}
%%%%%%%%%%%%%%%%%%%%%%%%%%%%%%%%%%%%%%%%%%%%%%%%%%

%%%%%%%%%%%%%%%%%%%%%%%%%%%%%%%%%%%%%%%%%%%%%%%%%%%%%%%%%%%%%%
\begin{algorithm}[htbp]
\caption{Pseudo-code of sampling a sector}
\label{alg: SampleSector algorithm} 
 \SetKwInOut{Input}{Input}
 \SetKwInOut{Output}{Output}
 \SetAlgoLined
 \SetKw{KwBreak}{break}
 \Input{Size of the central angle $\Delta \theta$, is\_fixed}
 \Output{A sector with $(\alpha,\beta)$}
 \tcp{Each sector is represented as $(\alpha,\beta)$}
 $\alpha \leftarrow\mathrm{randInt}(0,360) $\;
 \If{$!is\_fixed$}{
    $ \Delta \theta \leftarrow randInt(20,60)$\;
 }
 $\beta \leftarrow (\alpha + \Delta\theta) \% 360$\;
 
 sector = Sector($\alpha$,$\beta$);\tcp{Generate a sector area}
\end{algorithm}
%%%%%%%%%%%%%%%%%%%%%%%%%%%%%%%%%%%%%%%%%%%%%%%%%%%%%%%%%%%%

\subsubsection{Sector perturbation}
%%%%%%%%%%%%%%%%%%%%%%%%%%%%%%%%%%%%%%%%%%%%%%%%%%%%%%%%%
As Alg.~\ref{alg: SampleSector algorithm} shows, the sector perturbation strategy randomly generates a sector area $(\alpha, \beta)$ in a circular bin. The larger the central angle, the larger the sector area. Therefore, the larger the disturbance, the more circular items intersecting the area will be reassigned at each iteration. Especially the circular items are taken out and unassigned from the border to the center of the circular bin.
\begin{figure}[htbp]
\centering
\includegraphics[width=0.35\textwidth]{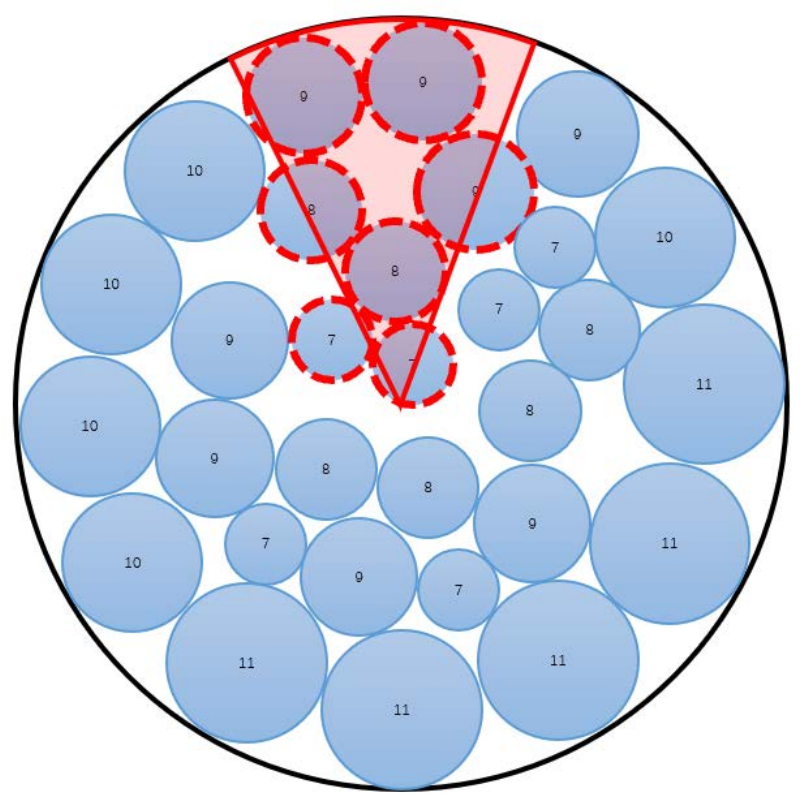}
\caption{An illustration of sector perturbation.} 
\label{fig: sector perturbation}
\end{figure}

 %%%%%%%%%%%%%%%%%%%%%%%%%%%%%%%%%%%%%%%%%%%%%%%%%%%%%%%%%%
\begin{algorithm}[htbp]
\caption{Pseudo-code of intersecting with the sector or circle}
\label{alg: Intersects}
\SetKwInOut{Input}{Input}
\SetKwInOut{Output}{Output}
\SetAlgoLined
\SetKw{KwBreak}{break}
\Input{Circle $C_i$, sector $S$ (or circle $C$), bin radius $R$}
\Output{True or false}\tcp{Returns if $C_i$ intersects the perturbation area $S$ (or $C$)}
\eIf{adopt the sector perturbation}
{
    $\alpha = S.\alpha$\;
    $\beta = S.\beta$\;
    \If{the center of $C_i$ is in sector area $S$}{
        return true;\tcp{Intersects with the sector area $S$}}
    \If{$C_i$ intersects with radii $r_{\alpha}$ or $r_{\beta}$ of $S$}{
        return true;\tcp{Intersects with the sector area $S$}
    }
    return false;\tcp{No intersects with the sector area $S$}
}
{
    \tcp{adopt the circle perturbation}
    \eIf{$(C_i.x-C.x)^{2} + (C_i.y-C.y)^{2} \le C_i.r + C.r$}{
    return true; \tcp{Intersects with circle area $C$}
    }{
    return false;\tcp{No intersects with circle area $C$}
    }
}
\end{algorithm}

Alg.~\ref{alg: Intersects} can determine whether a circular item $C_i$ intersects the selected circular (or sector) area. If a circular item $C_i$ intersects the chosen area, that is, the item with a red dotted border in Fig.~\ref{fig: circle perturbation} or Fig.~\ref{fig: sector perturbation}, which will be taken out from the bin and added to the unassigned circular set $circle\_ids$ in Alg.~\ref{alg: Generate New Solution} (line 5). However, as the sector area is not easy to express with mathematical formulas like the circle area, it is not intuitive to judge whether a circular item intersects with the sector area. Due to the sector is surrounded by two radii and the arc opposite the central angle. We turn it into two small subproblems: 1) whether the center of a circular item is in the sector area; 2) whether the circular item intersects the radii $r_{\alpha}$ or $r_{\beta}$ of the central angle. The former is judged by line 4, Alg.~\ref{alg: Intersects}, and the latter is implemented by line 7, Alg.~\ref{alg: Intersects}.
%%%%%%%%%%%%%%%%%%%%%%%%%%%%%%%%%%%%%%%%%%%%%%%%%%%%%%%%%%%
\begin{comment}
Alg.~\ref{alg: Intersect with line} determines whether a circular item intersects the given line, and the circular items crossing the border (radii) of the sector will be added to the unassigned set $circle\_ids$, as shown in line 5, Alg.~\ref{alg: Generate a Solution}.

%%%%%%%%%%%%%%%%%%%%%%%%%%%%%%%%%%%%%%%%%%%%%%%%%%%%%%%%%%%%%
\begin{algorithm}[htbp]
\caption{Intersect with Line}
\label{alg: Intersect with line} 
\SetKwInOut{Input}{Input}
\SetKwInOut{Output}{Output}
\SetAlgoLined
\SetKw{KwBreak}{break}
\Input{Circle $C_i$, angle, bin radius $R$}
\Output{True or false}\tcp{returns if $C_i$ Intersects with the line}
$\theta \leftarrow angle \times \pi / 180$\;
$C_0 \leftarrow R - C_i.x$\;
$C_1 \leftarrow R - C_i.y$\;
$a \leftarrow 1$\;
$b \leftarrow 2 \times C_0 \times cos\theta + 2 \times C_1 \times sin \theta$\;
$c \leftarrow C_0^2 + C_1^2 - C_i.r \times C_i.r $\;
$\Delta \leftarrow b^2 - 4ac$\;
$t_1 \leftarrow (-b + \sqrt{\Delta})/2$\;
$t_1 \leftarrow (-b - \sqrt{\Delta})/2$\;
\eIf{$t_1 \ge 0 \parallel t_2 \ge 0$}
{
    return true\;
}
{
return false;\tcp{No intersect with the line}}
\end{algorithm}
\end{comment}
%%%%%%%%%%%%%%%%%%%%%%%%%%%%%%%%%%%%%%%%%%%%%%%%%%%%%%%%%%%%%

%%%%%%%%%%%%%%%%%%%%%%%%%%%%%%%%%%%%%%%%%%%%%%%%%%%%%%%%%%%%%%
\begin{algorithm}[htbp]
\caption{Pseudo-code of generating a new solution}
\label{alg: Generate New Solution}
\SetKwInOut{Input}{Input}
\SetKwInOut{Output}{Output}
\SetAlgoLined
\SetKw{KwBreak}{break}
\Input{The old packing solution: $X$, bin radius: $R$ }
\Output{A new neighbor packing solution: $X^{'}$}
$K \leftarrow X.bins.size()$\;
\tcp{Select two bins randomly}
{$\left(k_1, k_2\right)\leftarrow$ random\_ints($2, \left\{1,\ldots,{K})\right\}$) \;
\tcp{Select a sector or circle area in the first bin using Algorithm~\ref{alg: SampleSector algorithm} or Algorithm~\ref{alg: SampleCircles-algorithm}}
$A_{1}\leftarrow$ Sample a sector$(\theta,is\_fixed)$; \tcp{or Sample a circle$(B_k,R)$;}

\tcp{Select an area in the second bin}
$A_{2}\leftarrow$ Sample a sector$(\theta,is\_fixed)$}; \tcp{ or Sample a circle$(B_k,R)$;}

$circle\_ids \leftarrow\bigcup_{j\in\left\{ 1,2\right\}}\{ i|\left\langle x_{i},y_{i},k_{j}\right\rangle \in X ~\bigwedge I_{ik_j}=1 
~\bigwedge \mathrm {intersects}(C_i,A_j,R)==True \}$; \tcp{See Algorithm~\ref{alg: Intersects}}

$bin\_ids \leftarrow {\{k_1,k_2\}}$\; 
$X^{'} \leftarrow TOA(circle\_ids,bin\_ids, R) $;\tcp{Generate a new solution}
\end{algorithm}
%%%%%%%%%%%%%%%%%%%%%%%%%%%%%%%%%%%%%%%%%%%%%%%%%%%%%%%%%%
\subsubsection{Generate Neighbor Solution}
\label{sec: subsubsec4.3.3}
As Alg.~\ref{alg: Generate New Solution} shows, a new packing solution $X^{'}$ is generated from the old packing solution $X$ by selecting two bins $B_{k_1}, B_{k_2}$ randomly and performing \textit{sector perturbation} or \textit{circle perturbation}. We randomly choose a sector area with equal angle size in each bin, and all items that intersect the sector area will be taken out, and their IDs will be added to set $circles\_ids$. $k_1$, $k_2$ will be added to set $bin\_ids$. The unassigned circular items will be reassigned with algorithm TOA. Then we will get a new neighbor packing solution $X^{'}$. 

At each iteration, two (or more) bins will be selected so that the unassigned circular items have more free space to be assigned. Even in the worst case, the algorithm will attempt to exchange the circular items in the two (or more) areas, ensuring that there will be some disturbance at each operation.

Besides, at the early stage, the sector area (i.e., $\Delta \theta$) can be set larger so that the new neighbor solution can be located far away from the current solution to speed up the search process and to avoid getting trapped at a local minimum solution. Once the temperature $f(X)$ gets low, the sector area will become smaller. The new solution will be generated nearby with the minor disturbance and focus on the local area.

\subsection{The Assignments of Parameters}
\label{sec:subsec4.4}
In the experiments, we find that different assignments of parameters are suitable for different problem scales. Therefore, to obtain a better solution in solving the packing problem in a broad scale, the parameter values should change along with the number of items, which can make the assignments of parameters dynamic and adaptive, such as the times of greedy search $t_{greedy}$:
\begin{equation}
    t_{greedy} \leftarrow \beta \times n,
    \label{eq:4.2}
\end{equation}
and the cool coefficient of the temperature $t_{cool}$:
\begin{equation} 
t_{cool} \leftarrow \frac{\alpha \times \sqrt{n} - 1}{\alpha \times \sqrt{n}}.
\label{eq:4.3}
\end{equation}
In this way, our algorithm is adaptive for the number of items, and the parameter space can be sampled much more efficiently. For example, if $n$ is small, we will get a quick cooling coefficient. As the number of items increases, the times of greedy search will become larger and get a slower cooling coefficient fit for big-scale packing instances. The difficulty of the problem becomes higher as the number of items becomes larger, indicating more solution space to be explored.

\subsection{The Overall ASA-GS Algorithm}
\label{sec: subsec4.5}
The workflow of ASA-GS is provided in Alg.~\ref{alg: ASA-GS}.
 Firstly, it is necessary to initialize the annealing parameters and attain a feasible packing pattern with the initial solution as shown in subsection~\ref{sec:subse4.2}. Then, it will select one of the perturbation methods between sector perturbation and circle perturbation as well as generate a new neighbor packing solution by Alg.~\ref{alg: Generate New Solution}. After that, it will compute the energy function and utilize the greedy search technique based on the simulated annealing to decide whether accept the new solution. Finally, it updates the parameters such as the cooling coefficient of the temperature with $t_{current} = t_{current} \times t_{cool}$, and the acceptance probability by Eq.~(\ref{eq:4.4}). The process will execute until the terminal criterion such as the current temperature $t_{current}$ is below the threshold $t_{end}$, or the number of iterations $i$ exceeds the given value $N$.
 
The key concept of greedy search can be described as follows: take a new neighbor packing $X^{'}(i.e. X_1^{'})$ as the best packing $X$ when $dE \le 0 (i.e. f(X^{'}) \le f(X))$, and go to the next step. Otherwise the algorithm continues to generate the next new neighbor packing $X_2^{'}$, and takes it as the best packing $X$ when $f(X_2^{'}) \le f(X)$, then goes to the next step. Otherwise this step will continue to be executed until attaining a better packing solution or has generated $t_{greedy}-th$ new neighbor packing $X_{t_{greedy}}^{'}$. The latter will generate $t_{greedy}$ neighbor packing solutions $X_1^{'}$, $X_2^{'}$\dots, $X_{t_{greedy}}^{'}$ while they are all worse than the original packing solution $X$. In such case, it will accept the best new packing $X^{'}_{best}$ among the $t_{greedy}$ neighbor packing solutions generated with probability $p$.

\begin{equation} 
p \leftarrow e^{-(f(X^{'}) - f(X))/t_{current} \times log(n/2) }.
\label{eq:4.4}
\end{equation}

Obviously, the quality of the best neighbor solution $X^{'}_{best}$ will vary from low to high with the times of greedy search increases so that the new solution can jump to a better solution space with high probability. $f(X^{'}_{best})$ is defined by Eq. (\ref{eq:4.5}):
\begin{equation}
    f(X^{'}_{best})=min(f(X^{'}_{1}),f(X^{'}_{2}),...,f(X^{'}_{G}),...,f(X^{'}_{t_{greedy}})).
    \label{eq:4.5}
\end{equation}

The ASA-GS algorithm can achieve faster convergence and improve the quality-time trade-off by utilizing the greedy search technique. As the experimental results show, the solutions produced by the ASA-GS algorithm are very competitive. %fairly satisfactory.

%% file: Sec5-Exp.tex
For experiments, we evaluate and analyze the competency and performance of the proposed algorithms, TOA and ASA-GS. 
%Various experiments were conducted to assess their performance for CBPP-CI. 
We implemented the algorithms using Visual C++ programming language. All results were generated by setting parameters as $N=2\times10^{6}$, $\alpha = 0.9$, $\beta = 0.08$, $t_{start} = 0.1$, $t_{end} = 10^{-4}$, and obtained using a computer equipped with an Intel(R) Core(TM) i7-10710U CPU @ 1.10GHz 1.61Hz. 

As the CBPP-CI is a new problem, there are no available benchmark instances. Referring to the pioneering work of square bin packing problem with circular items (SBPP-CI)~\citep{HE2021105140}. We choose two categories of benchmarks for the single circle packing problem (SCPP) on the packomonia website and build two sets of new benchmark instances based on them for the CBPP-CI. 

%%%%%%%%%%%%%%%%%%%%%%%%%%%%%%%%%%%%%%%%%%%%%%%%%%%%%%%%%%%%%
\begin{table*}[pos=htbp]
\caption{Experimental results on the fixed
benchmarks with circular bins  for $r_i = i$.}
\label{tab:one}
\begin{tabular*}{\tblwidth}{@{}LLLLLLLLLLLL@{}}
\toprule
$n_{0}$ & $n$ & $Alg.$ & $Bin_0$ & bin 1 & bin 2 & bin 3 & bin 4 & bin 5 & bin 6 & $F$ & $F_{A}-F_{T}$\\
\midrule
{8} & {40}& {ASA-GS} & 0.78 & 0.84 & 0.80 & 0.74 & 0.74 & 0.71 & 0.03 & $-5.19$ & {0.19}\\
%\cline{5-11} 
 &  & TOA && 0.81 & 0.74 & 0.72 & 0.72 & 0.69 & 0.19 & $-5.38$ & \\
%\hline  
{9} & {45} & ASA-GS & 0.79 & 0.83 & 0.80 & 0.77 & 0.76 & 0.70 & - & -4.87 & {0.47}\\
%\cline{5-11}  
 &  & TOA && 0.81 & 0.75 & 0.74 & 0.71 & 0.69 & 0.15 & -5.34 & \\
%\hline  
{10} & {50} & ASA-GS  &0.80& 0.84 & 0.79 & 0.79 & 0.79 & 0.79 & - & -4.95 & {0.45}\\
%\cline{5-11}  
 &  & TOA && 0.81 & 0.79 & 0.74 & 0.74 & 0.70 & 0.21 & -5.40 & \\
%\hline  
{11} & {55} & ASA-GS &0.80& 0.84 & 0.81 & 0.81 & 0.77 & 0.77 & 0.07 & -5.23 & {0.29}\\
%\cline{5-11}  
 &  & TOA && 0.83 & 0.74 & 0.72 & 0.71 & 0.70 & 0.35 & -5.52 & \\
%%\hline  
{12} & {60} & ASA-GS &0.80& 0.85 & 0.80 & 0.79 & 0.77 & 0.76 & 0.06 & -5.21 & {0.21}\\
%\cline{5-11}  
 &  & TOA && 0.82 & 0.79 & 0.77 & 0.73 & 0.68 & 0.24 & -5.42 & \\
%\hline  
{13} & {65} & ASA-GS &0.81& 0.84 & 0.82 & 0.81 & 0.78 & 0.76 & 0.10 & -5.26 & {0.16}\\
%\cline{5-11}  
 &  & TOA && 0.84 & 0.81 & 0.75 & 0.75 & 0.72 & 0.26 & -5.42 & \\
%\hline  
{14} & {70} &ASA-GS&0.81 & 0.85 & 0.80 & 0.79 & 0.79 & 0.78 & 0.12 & -5.27 & {0.11}\\
%\cline{5-11}  
 &  & TOA & & 0.86 & 0.80 & 0.77 & 0.74 & 0.72 & 0.24 & -5.38 & \\
%\hline  
{15} & {75} & ASA-GS& 0.82 & 0.85 & 0.82 & 0.81 & 0.79 & 0.77 & 0.08 & -5.23 & {0.19}\\
%\cline{5-11}  
 &  & TOA && 0.84 & 0.79 & 0.75 & 0.74 & 0.72 & 0.26 & -5.42 & \\
%\hline  
{16} & {80} & ASA-GS&0.83 & 0.85 & 0.82 & 0.82 & 0.78 & 0.78 & 0.11 & -5.26 & {0.13}\\
%\cline{5-11}  
 &  & TOA& & 0.86 & 0.82 & 0.77 & 0.74 & 0.70 & 0.25 & -5.39 & \\
%\hline  
{17} & {85} & ASA-GS & 0.83& 0.86 & 0.83 & 0.81 & 0.79 & 0.76 & 0.11 & -5.25 & {0.12}\\
%\cline{5-11}  
 &  & TOA& & 0.86 & 0.84 & 0.75 & 0.74 & 0.74 & 0.23 & -5.37 & \\
%\hline  
{18} & {90} & ASA-GS &0.83& 0.86 & 0.83 & 0.80 & 0.79 & 0.78 & 0.14 & -5.28 & {0.11}\\
%\cline{5-11}  
 &  & TOA & & 0.86 & 0.83 & 0.76 & 0.75 & 0.75 & 0.25 & -5.39 & \\
%\hline  
{19} & {95} & ASA-GS&0.84 & 0.86 & 0.83 & 0.80 & 0.79 & 0.77 & 0.15 & -5.29 & {0.12}\\
%\cline{5-11}  
 &  & TOA & & 0.86 & 0.82 & 0.77 & 0.75 & 0.73 & 0.27 & -5.41 & \\
%\hline  
{20} & {100} & ASA-GS&0.84 & 0.87 & 0.83 & 0.80 & 0.80 & 0.77 & 0.13 & -5.26 & {0.12}\\
%\cline{5-11}  
 &  & TOA && 0.86 & 0.83 & 0.77 & 0.76 & 0.75 & 0.24 & -5.38 & \\
\bottomrule
\end{tabular*}
\end{table*}
%%%%%%%%%%%%%%%%%%%%%%%%%%%%%%%%%%%%%%%%%%%%%%%%%%%%%%%%%%%%%%

The generated instances consist of strong heterogeneous $r_i=i$ (i.e., the circle radii vary widely), and $r_{i}=\sqrt{i}$ for weakly heterogeneous instances. For each category, we produce fixed and random instances. We first choose instances from the packomonia website for SCPP to generate our instances. Each circular bin's best-known solution found in \cite{packomania} ranges from 8 to 20 from the circular bin benchmarks. The fixed set of benchmarks contains exactly five copies of each circle instance, and for the random benchmarks instances, it contains a random copy of each circular item that ranges from $2-10$ from the same benchmarks. We fix the circular bin size from the best solution found on the packomonia website.

In the computational tables, we list 52 generated instances from the two categories of benchmarks (fixed and rand). For each instance in the Tables ( \ref{tab:one}, \ref{tab:two}, \ref{tab:three} and \ref{tab:four}), we have results for two algorithms: ASA-GS and TOA. Column $n_{0}$ represents the original index number of the circle set for each instance, column $n$ represents the actual number of replicated circles in the CBPP-CI instance. The third column (i.e., Alg.) represents the two algorithms. Column $Bin_0$ is only for the fixed benchmarks representing the reference value indexed from \cite{packomania} for the state-of-the-art results. Columns $5^{th}$ to $10^{th}$ denote the density (bin occupancy rate) for each bin. Lastly, the $F$ and $F_A-F_T$ columns represent the actual measure value achieved for each algorithm and relative improvement of ASA-GS over TOA. 

\subsection{Comparison on $r=i$}
\label{sec:subsec5.1}
Here $r=i$ is a benchmark instance that has a wide variation of circle sizes. In this set of benchmarks, we execute ASA-GS and TOA algorithms for comparison. We select instances that range from 8 to 20 for both fixed and random setup from the benchmark. Table \ref{tab:one} displays the computational results of fixed benchmarks while Table \ref{tab:two} displays for random benchmarks.

%%%%%%%%%%%%%%%%%%%%%%%%%%%%%%%%%%%%%%%%%%%%%%%%%%%%%%%%%%%%%
\begin{table*}[pos=htbp]
\caption{Experimental results on the random
benchmarks with circular bins for $ r_i = i$.}
\label{tab:two}
\begin{tabular*}{\tblwidth}{@{}LLLLLLLLLLL@{}}
\toprule
$n_{0}$ & $n$ & $Alg.$ & bin 1 & bin 2 & bin 3 & bin 4 & bin 5 & bin 6 & $F$ & $F_{A}-F_{T}$\\
\midrule
{8} & {35} & ASA-GS & 0.84 & 0.81 & 0.71 & - & - & - & -2.87 & {0.36}\\
%\cline{3-10}  
 &  & TOA & 0.84 & 0.77 & 0.68 & 0.07 & - & - & -3.23& \\
%%\hline  
{9} & {44} & ASA-GS & 0.84 & 0.80 & 0.80 & 0.70 & - & - & -3.86 & {0.41}\\
%\cline{3-10} 
 &  & TOA & 0.82 & 0.75 & 0.75 & 0.73 & 0.09 & - & -4.27 & \\
%%\hline  
{10} & {48} & ASA-GS & 0.84 & 0.84 & 0.79 & 0.70 & - & - & -3.86 & {0.39}\\
%\cline{3-10} 
 &  & TOA & 0.83 & 0.80 & 0.74 & 0.72 & 0.08 & - & -4.25 & \\
%%\hline  
{11} & {52} & ASA-GS & 0.85 & 0.84 & 0.82 & 0.71 & - & - & -3.86 & {0.39}\\
%\cline{3-10}  
 &  & TOA & 0.85 & 0.80 & 0.74 & 0.73 & 0.10 & - & -4.25 & \\
%%\hline  
{12} & {59} & ASA-GS & 0.85 & 0.82 & 0.81 & 0.72 & - & - & -3.87 & {0.37}\\
%\cline{3-10}  
 &  & TOA & 0.85 & 0.80 & 0.74 & 0.72 & 0.09 & - & -4.24 & \\
%%\hline  
{13} & {64} & ASA-GS & 0.85 & 0.83 & 0.83 & 0.73 & - & - & -3.88 & {0.35}\\
%\cline{3-10}   
 &  & TOA & 0.85 & 0.80 & 0.77 & 0.74 & 0.08 & - & -4.23 & \\
%%\hline  
{14} & {67} & ASA-GS & 0.85 & 0.82 & 0.79 & 0.79 & 0.16 & - & -4.31 & {0.07}\\
%\cline{3-10}  
 &  & TOA & 0.87 & 0.81 & 0.77 & 0.72 & 0.25 & - & -4.38 & \\
%%\hline  
{15} & {73} & ASA-GS & 0.87 & 0.82 & 0.78 & 0.65 & - & - & -3.78 & {0.10}\\
%\cline{3-10}  
 &  & TOA & 0.85 & 0.80 & 0.74 & 0.73 & - & - & -3.88 & \\
%%\hline  
{16} & {79} & ASA-GS & 0.86 & 0.83 & 0.81 & 0.80 & 0.73 & - & -4.87 & {0.39}\\
%\cline{3-10}  
 &  & TOA & 0.86 & 0.82 & 0.77 & 0.76 & 0.70 & 0.12 & -5.26 & \\
%%\hline  
{17} & {84} & ASA-GS & 0.86 & 0.84 & 0.80 & 0.78 & 0.10 & - &  -4.24 & {0.08}\\
%\cline{3-10}  
 &  & TOA & 0.86 & 0.82 & 0.78 & 0.74 & 0.18 & - & -4.32 & \\
%%\hline  
{18} & {87} & ASA-GS & 0.87 & 0.83 & 0.80 & 0.80 & 0.71 & - & -4.84 & {0.44}\\
%\cline{3-10} 
 &  & TOA & 0.85 & 0.82 & 0.75 & 0.74 & 0.72 & 0.13 & -5.28 & \\
%%\hline  
{19} & {92} & ASA-GS & 0.87 & 0.84 & 0.82 & 0.80 &  0.06 & - & -4.19 & {0.09}\\
%\cline{3-10} 
 &  & TOA & 0.86 & 0.83 & 0.80 & 0.76 & 0.14 & - &-4.28 & \\
%%\hline  
{20} & {97} & ASA-GS & 0.87 & 0.84 & 0.81 & 0.74 & - & - & -3.87 & {0.40}\\
%\cline{3-10} 
 &  & TOA & 0.86 & 0.82 & 0.74 & 0.71 & 0.13 & - & -4.27 & \\
\bottomrule
\end{tabular*}
\hfill{}
\end{table*}
%%%%%%%%%%%%%%%%%%%%%%%%%%%%%%%%%%%%%%%%%%%%%%%%%%%%%%%%%%%%%

\begin{figure*}[pos=htbp]
\centering
\includegraphics[width = 0.833\textwidth]{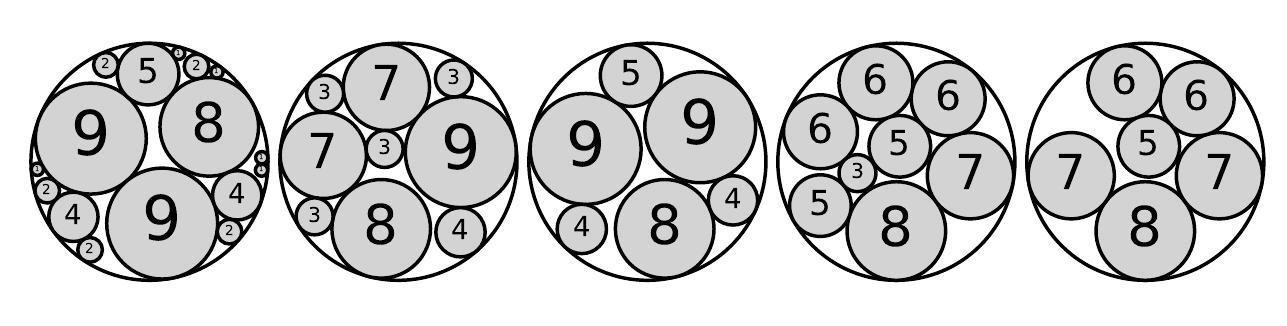}
\includegraphics[width = 1\textwidth]{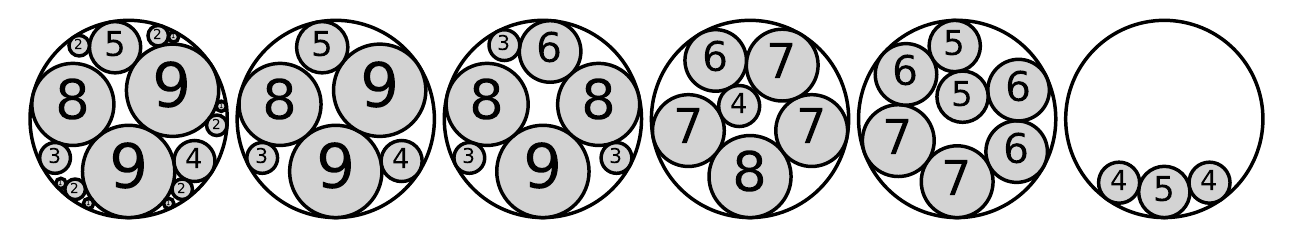}
\caption{Packing layouts generated by ASA-GS (top) and TOA (bottom) for the fixed benchmark $r = i$ with $9\times5$
circles.}
\label{fig:9 45}
\end{figure*}

In Table~\ref{tab:one} we can notice that the objective value of ASA-GS is better than TOA on all the instances, and in addition, we can also observe one lesser bin occupancy rate. for instance, $n_0=9 ~\& ~n=45$ and $n_0=10 ~\&~ n=45$, i.e., ASA-GS uses five bins to pack 45 circles while TOA uses six bins to load the same set of circular items, for a diagrammatic representation of the packing layout when $n_0=9~\&~n=45$ (See Fig.~\ref{fig:9 45}) and when $n_0=10~\&~n=50$, we can also notice that ASA-GS packs 50 circles in 5 bins while TOA uses six bins for the same set of circular items. For the fixed benchmarks, ASA-GS has an average of $21\%$ improvement.

%%%%%%%%%%%%%%%%%%%%%%%%%%%%%%%%%%%%%%%%%%%%%%%%%%%%%%%%%%%%%
%%%%%%%%%%%%%%%%%%%%%%%%%%%%%%%%%%%%%%%%%%%%%%%%%%%%%%%%%%%%%%%

%%%%%%%%%%%%%%%%%%%%%%%%%%%%%%%%%%%%%%%%%%%%%%%%%%%%%%%%%%%%%%%
\begin{figure*}[pos=htbp]
\centering
\includegraphics[width = 0.67\textwidth]{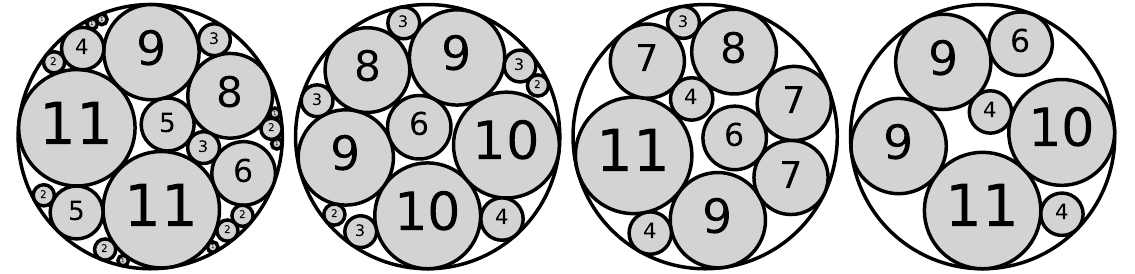}
\includegraphics[width =0.833 \textwidth]{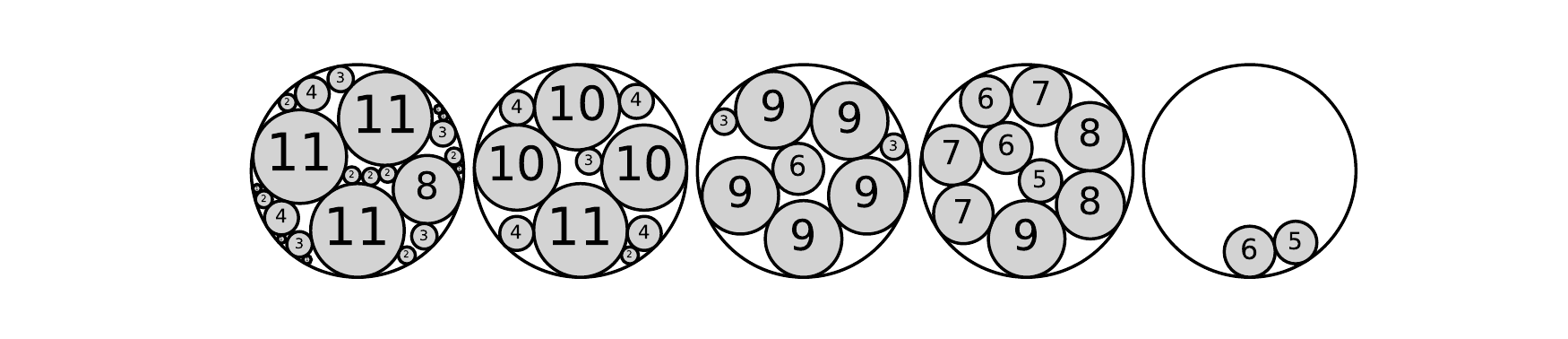}
\caption{Packing layouts generated by ASA-GS (top) and TOA (bottom) for the random benchmark $r_i = i$ with $11-52$
circles.}
\label{fig:11 52}
\end{figure*}
%%%%%%%%%%%%%%%%%%%%%%%%%%%%%%%%%%%%%%%%%%%%%%%%%%%%%%%%%%%%%%

For the random benchmarks, we can also observe that in all the instances, ASA-GS returns a feasible solution compared to TOA with an overall average improvement of $30\%$ for $r=i$ benchmarks in Table~\ref{tab:two}. We show the packing layout when $n_0 = 11$~\&~$ n = 52$  in Fig.~\ref{fig:11 52} for the random benchmarks. 

\subsection{Comparison on $r_{i}=\sqrt{i}$}
\label{sec:subsec5.2}
Here $r_{i}=\sqrt{i}$ has a smaller variation of the circle's radii. Similarly, we test the two algorithms on this set of benchmarks ranging from $8-20$ instances. Table \ref{tab:three} and \ref{tab:four} represent fixed and random benchmarks respectively.

%%%%%%%%%%%%%%%%%%%%%%%%%%%%%%%%%%%%%%%%%%%%%%%%%%%%%%%%%%%%%
\begin{table*}[pos=htbp]
\caption{Experimental results on the fixed
benchmarks with circular bins for $r_i = \sqrt{i}$.}
\label{tab:three}
\begin{tabular*}{\tblwidth}{@{}LLLLLLLLLL@{}}
%\hline  
\toprule
$n_{0}$ & $n$ & $Alg.$ & $Bin_0$ & bin 1 & bin 2 & bin 3 & bin 4 & $F$ & $F_{A}-F_{T}$\\
%\hline  
\midrule
{8} & {40} & ASA-GS & 0.76 & 0.84 & 0.78 & 0.76 & 0.39 & -3.55 & {0.09}\\
%\cline{5-9} 
 &  & TOA && 0.81 & 0.76 & 0.76 & 0.45 & -3.64 & \\
%%\hline   
{9} & {45} & ASA-GS &0.77& 0.83 & 0.81 & 0.76 & 0.37 & -3.54 & {0.12}\\
%\cline{5-9}
 &  & TOA & & 0.83 & 0.75 & 0.70 & 0.49 & -3.66 & \\
%%\hline   
{10} & {50} & ASA-GS &0.80 & 0.83 & 0.80 & 0.78 & 0.39 & -3.56 & {0.09}\\
%\cline{5-9} 
 &  & TOA & & 0.83 & 0.75 & 0.74 & 0.48 & -3.65 & \\
%%\hline   
{11} & {55} & ASA-GS &0.81& 0.86 & 0.80 & 0.79 & 0.37 & -3.51 & {0.06}\\
%\cline{5-9} 
 &  & TOA && 0.85 & 0.78 & 0.74 & 0.42 & -3.57 & \\
%%\hline   
{12} & {60} & ASA-GS &0.81& 0.85 & 0.81 & 0.80 & 0.32 & -3.47 & {0.09}\\
%\cline{5-9} 
 &  & TOA && 0.86 & 0.76 & 0.75 & 0.42 & -3.56 & \\
%%\hline   
{13} & {65} & ASA-GS &0.82& 0.85 & 0.81 & 0.79 & 0.37 & -3.52 & {0.07}\\
%\cline{5-9} 
 &  & TOA && 0.85 & 0.78 & 0.74 & 0.44 & -3.59 & \\
%\hline  
{14} & {70} & ASA-GS &0.82& 0.86 & 0.80 & 0.78 & 0.37 & -3.51 & {0.13}\\
%\cline{5-9} 
 &  & TOA && 0.83 & 0.76 & 0.75 & 0.47 & -3.64 & \\
%\hline  
{15} & {75} & ASA-GS &0.82& 0.85 & 0.81 & 0.79 & 0.38 & -3.53 & {0.08}\\
%\cline{5-9} 
 &  & TOA && 0.86 & 0.76 & 0.75 & 0.47 & -3.61 & \\
%\hline  
{16} & {80} & ASA-GS &0.83& 0.86 & 0.80 & 0.80 & 0.39 & -3.53 & {0.06}\\
%\cline{5-9} 
 &  & TOA && 0.85 & 0.79 & 0.76 & 0.44 & -3.59 & \\
%\hline  
{17} & {85} & ASA-GS &0.83& 0.86 & 0.80 & 0.80 & 0.39 & -3.53 & {0.07}\\
%\cline{5-9} 
 &  & TOA && 0.85 & 0.78 & 0.77 & 0.45 & -3.60 & \\
%\hline  
{18} & {90} & ASA-GS &0.84& 0.87 & 0.81 & 0.80 & 0.38 & -3.51 & {0.06}\\
%\cline{5-9} 
 &  & TOA &&0.87 & 0.79 & 0.77 & 0.44 & -3.57 & \\
%\hline  
{19} & {95} & ASA-GS &0.84& 0.87 & 0.81 & 0.79 & 0.40 & -3.53 & {0.07}\\
%\cline{5-9}
 &  & TOA && 0.86 & 0.78 & 0.77 & 0.46 & -3.60 & \\
%\hline  
{20} & {100} & ASA-GS &0.84 & 0.86 & 0.82 & 0.80 & 0.40 & -3.54 & {0.06}\\
%\cline{5-9} 
 &  & TOA && 0.87 & 0.78 & 0.75 & 0.47 & -3.60 & \\
\bottomrule
\end{tabular*}
\end{table*}
%%%%%%%%%%%%%%%%%%%%%%%%%%%%%%%%%%%%%%%%%%%%%%%%%%%%%%%%%%

\begin{figure*}[pos=htbp]
\centering
\includegraphics[width = 0.67\textwidth]{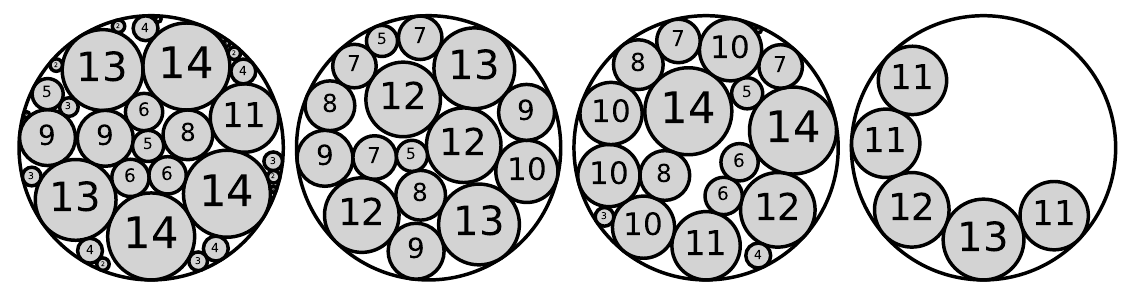}
\includegraphics[width = 0.67\textwidth]{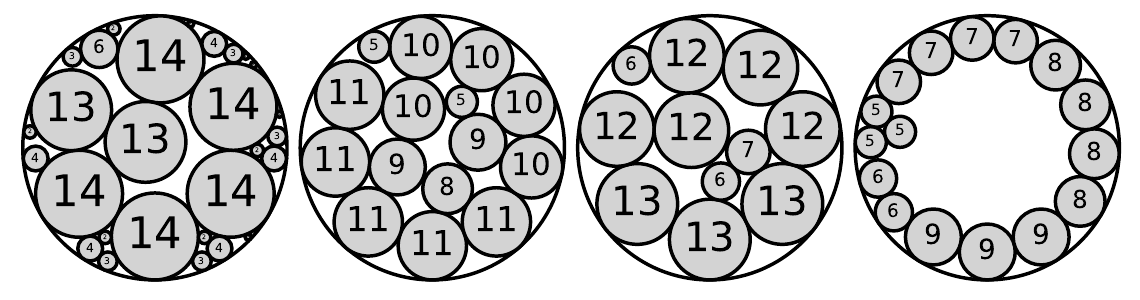}
\caption{Packing layouts generated by ASA-GS (top) and TOA (bottom) for the fixed benchmark $r_i = \sqrt{i}$ with $14\times5$ circles.}
\label{fig:14 70}
\end{figure*}

Table \ref{tab:three} is for the fixed benchmarks, from which we can notice that ASA-GS outperforms TOA in all the instances with an average improvement of 8\%.
We show the significant improvements of the packing layout in Fig.~\ref{fig:14 70}.  

\begin{figure*}[htbp]
\centering
\includegraphics[width = 0.5\textwidth]{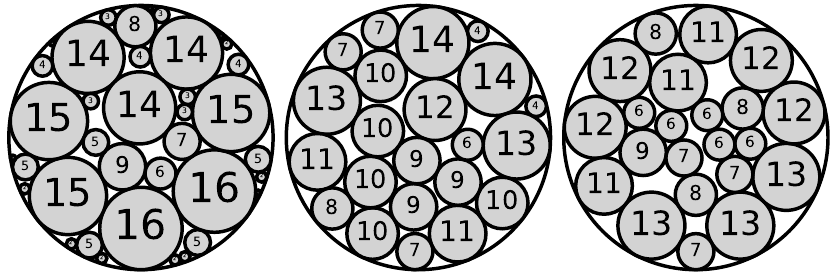}
\includegraphics[width = 0.67\textwidth]{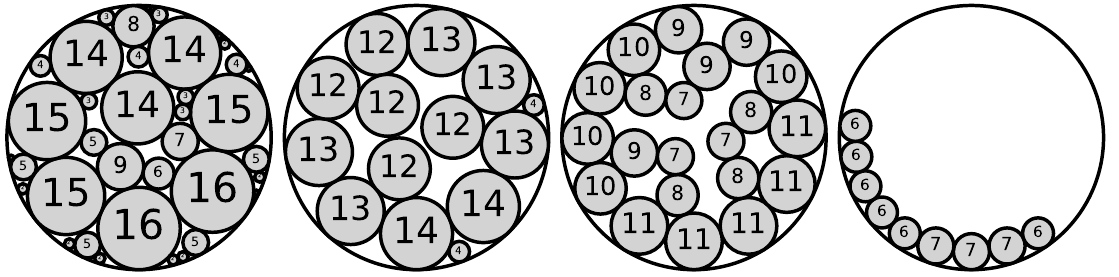}
\caption{Packing layouts generated by ASA-GS (top) and TOA (bottom) for the random benchmark $r_i = \sqrt{i}$ with $16-81$
circles.}
\label{fig:16 81}
\end{figure*}

We can notice that ASA-GS packs the items with lesser bins than TOA for the random benchmarks in Table~\ref{tab:four}. We can observe that when $n_0=8$ (or 9, 10, 12, 13, 15, 16, 18, 19), ASA-GS uses fewer bins than TOA. We use $n_0=16~\&~n = 81$ to demonstrate the packing layout for this benchmark in Fig.~\ref{fig:16 81}. For these random benchmarks, we can notice an overall improvement of $26\%$.

%%%%%%%%%%%%%%%%%%%%%%%%%%%%%%%%%%%%%%%%%%%%%%%%%%%%%%
\begin{table*}[pos=htbp]
\centering
\caption{Experimental results on the random
benchmarks with circular bins for $r_i = \sqrt{i}$.}
\label{tab:four}
\begin{tabular*}{\tblwidth}{@{}LLLLLLLLL@{}}
%%\hline  
\toprule
$n_{0}$ & $n$ & $Alg.$ & bin 1 & bin 2 & bin 3 & bin 4 & $F$ & $F_{A}-F_{T}$\\
%%\hline  
\midrule
{8} & {29} & ASA-GS & 0.83 & 0.79 & - & - & -1.96 & {0.29}\\
%\cline{3-8} 
 &  & TOA & 0.82 & 0.72 & 0.07 & - & -2.25 & \\
%\hline  
{9} & {36} & ASA-GS & 0.84 & 0.75 & - & - & -1.91 & {0.34}\\
%\cline{3-8} 
 &  & TOA & 0.80 & 0.74 & 0.05 & - & -2.25 & \\
%\hline  
{10} & {51} & ASA-GS & 0.86 & 0.81 & 0.73 & - & -2.87 & {0.34}\\
%\cline{3-8} 
 &  & TOA & 0.84 & 0.76 & 0.76 & 0.05 & -3.21 & \\
%\hline  
{11} & {56} & ASA-GS & 0.85 & 0.80 & 0.68 & - & -2.83 & {0.06}\\
%\cline{3-8} 
 &  & TOA & 0.85 & 0.75 & 0.74 & - & -2.89 & \\
%\hline  
{12} & {61} & ASA-GS & 0.86 & 0.82 & 0.73 & - & -2.87 & {0.37}\\
%\cline{3-8} 
 &  & TOA & 0.83 & 0.76 & 0.75 & 0.07 & -3.24 & \\
%\hline  
{13} & {63} & ASA-GS & 0.86 & 0.82 & 0.74 & - & -2.88 & {0.33}\\
%\cline{3-8} 
 &  & TOA & 0.84 & 0.79 & 0.74 & 0.05 & -3.21 & \\
%\hline  
{14} & {66} & ASA-GS & 0.86 & 0.81 & 0.65 & - & -2.79 & {0.12}\\
%\cline{3-8} 
 &  & TOA & 0.83 & 0.75 & 0.74 & - & -2.91 & \\
%\hline  
{15} & {77} & ASA-GS & 0.86 & 0.83 & 0.73 & - & -2.87 & {0.38}\\
%\cline{3-8} 
 &  & TOA & 0.85 & 0.76 & 0.72 & 0.10 & -3.25 & \\
%\hline  
{16} & {81} & ASA-GS & 0.86 & 0.81 & 0.78 & -& -2.92 & {0.36}\\
%\cline{3-8} 
 &  & TOA & 0.86 & 0.76 & 0.7 & 0.14 & -3.28 & \\
%\hline  
{17} & {83} & ASA-GS & 0.86 & 0.81 & 0.79 & 0.07 & -3.21 & {0.12}\\
%\cline{3-8} 
 &  & TOA & 0.85 & 0.76 & 0.74 & 0.18 & -3.33 & \\
%\hline  
{18} & {89} & ASA-GS & 0.86 & 0.84 & 0.75 & - & -2.89 & {0.30}\\
%\cline{3-8} 
 &  & TOA & 0.87 & 0.77 & 0.76 & 0.06 & -3.19 & \\
%\hline  
{19} & {93} & ASA-GS & 0.87 & 0.81 & 0.77 & - & -2.90 & {0.31}\\
%\cline{3-8} 
 &  & TOA & 0.86 & 0.80 & 0.74 & 0.07 & -3.21 & \\
%\hline  
{20} & {97} & ASA-GS & 0.85 & 0.81 & 0.80 & 0.05 & -3.2 & {0.09}\\
%\cline{3-8} 
 &  & TOA & 0.85 & 0.78 & 0.75 & 0.14 & -3.29 & \\
\bottomrule
\end{tabular*}
\end{table*}
%%%%%%%%%%%%%%%%%%%%%%%%%%%%%%%%%%%%%%%%%%%%%%%%%%%%%%%%%%%%%

%%%%%%%%%%%%%%%%%%%%%%%%%%%%%%%%%%%%%%%%%%%%%%%%%%%%%%%%%%%%%%%
%%%%%%%%%%%%%%%%%%%%%%%%%%%%%%%%%%%%%%%%%%%%%%%%%%%%%%%%%%%%%

%%%%%%%%%%%%%%%%%%%%%%%%%%%%%%%%%%%%%%%%%%%%%%%%%%%%%%%%%%%%%
%%%%%%%%%%%%%%%%%%%%%%%%%%%%%%%%%%%%%%%%%%%%%%%%%%%%%%%%%%%%%

 \subsection{Further Analysis}
 \label{sec:subsec5.3}

 %%%%%%%%%%%%%%%%%%%%%%%%%%%%%%%%%%%%%%%%%%%%%%%%%%%%%%%%%%%%%
 \begin{table}[pos=htbp!]
 \centering
\caption  {T--test statistical analysis.}\label{tab:pvalue}
\begin{tabular*}{\tblwidth}{@{}LLL@{}}
\toprule
Group & Table & p--value\\
\midrule
{$r_i = i$} & Table \ref{tab:one} & 0.0000662229 \\
%\cline{2-3}  
   & Table \ref{tab:two} & 0.0000107770  \\
%\hline  
{$r_{i}=\sqrt{i}$} & Table \ref{tab:three} & 0.0000000252 \\
%\cline{2-3}
 &  Table \ref{tab:four} & 0.0000037048 \\
\bottomrule
\end{tabular*}
\end{table}
%%%%%%%%%%%%%%%%%%%%%%%%%%%%%%%%%%%%%%%%%%%%%%%%%%%%%%%%
 Since our solution methods are stochastic, we further analyze and assess the two proposed algorithms' significance comparison by using a T-tail statistical hypothesis test on $H_0$: $\mu_T = \mu_A$. $H_0$ denotes the null hypothesis, which equates to no difference between the results returned by TOA and ASA-GS.  We apply the commonly used $\alpha  = 0.05$ as our thresh-hold value. For each table we generated the p-value and compared with the $\alpha  = 0.05$ value as shown in Table~\ref{tab:pvalue}. We reject the null hypothesis from the generated results and claim with a  confidence interval (CI) of 95\% that our proposed algorithms are statistically distinct.
 
 To further demonstrate the typical performance pattern of the two algorithms, we illustrate the performance comparisons of the two algorithms. 

%%%%%%%%%%%%%%%%%%%%%%%%%%%%%%%%%%%%%%%%%%%%%%%%%%%%%%%%%%%%%
 \begin{figure*}[pos=htbp]
    \centering
    \includegraphics[width = 1\textwidth]{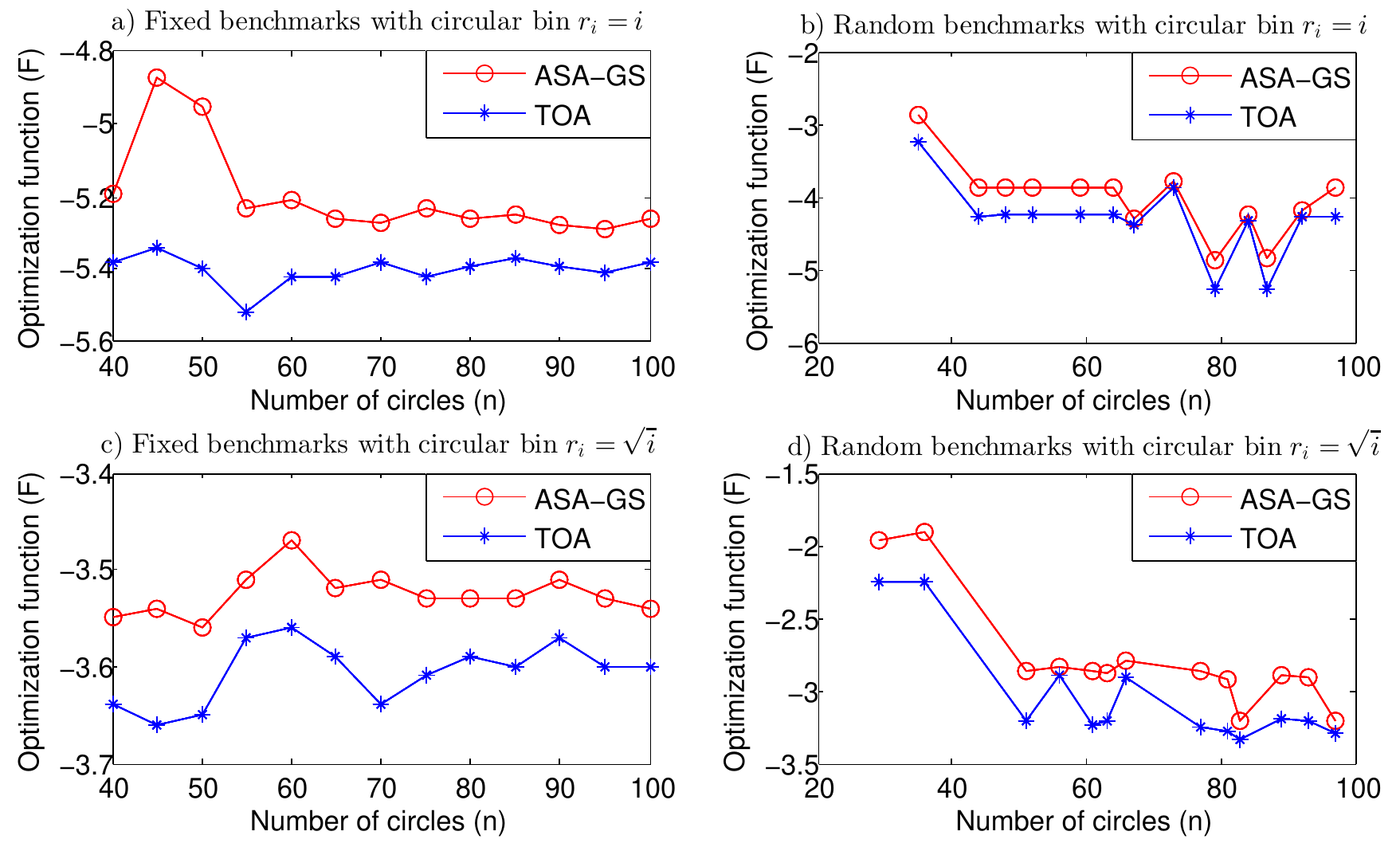}
    \caption{ASA-GS versus TOA.}
    \label{fig: result ASA-GS versus TOA}
\end{figure*}
%%%%%%%%%%%%%%%%%%%%%%%%%%%%%%%%%%%%%%%%%%%%%%%%%%%%%%%%%%%%%

 In Fig.~\ref{fig: result ASA-GS versus TOA}, the \emph{Y}-axis represents the optimization function while the \emph{X}-axis represents the number of circles ($n$). A red (blue) line presents ASA-GS (TOA). We can notice a distinct variation of ASA-GS and TOA lines that do not intersect, indicating that the ASA-GS completely outperforms the base TOA on all instances. 
 
 Lastly, we record the runtimes for $r=i$ and $r_{i}=\sqrt{i}$ benchmarks as shown in Table \ref{tab:Runtime}. The execution time of TOA is in micro-seconds while ASA-GS takes less than 200 seconds. In summary, the performance clearly shows ASA-GS efficiency outperforms TOA in a reasonable amount of time in all the instances. And in some instances, we can notice a reduction in the number of bins used. Moreover, Table~\ref{tab:two} and Table~\ref{tab:four} show that the density of $bin 1$ and $bin 2$ is usually greater than that of $Bin_0$. It indicates that the packing density of ASA-GS on the top few bins is much higher than the best packing results for SCPP on the packomonia website, inferring the high quality of our solution for the CBPP-CI.
 
 %%%%%%%%%%%%%%%%%%%%%%%%%%%%%%%%%%%%%%%%%%%%%%%%%%%%%%%%%%%%%
\begin{table*}[pos=t]%htbp]
\caption {Runtimes for ASA-GS execution on all benchmarks.}
\label{tab:Runtime}
\begin{tabular*}{\tblwidth}{@{}LLLLL|LLLL@{}}
\toprule
 & \multicolumn{4}{c|}{$r_i = i$} & \multicolumn{4}{c}{$r_{i}=\sqrt{i} $}\\
\midrule
 & \multicolumn{2}{c}{fixed} & \multicolumn{2}{c|}{random} & \multicolumn{2}{c}{fixed} & \multicolumn{2}{c}{random}\\
%%\hline  
\midrule
 $n_0$ & $n$ & $t$ & $n$ & $t$ & $n$ & $t$ & $n$ & $t$\\
%%\hline  
\midrule
8 & 40 & 12 & 35 & 34 & 40 & 24 & 29 & 53\\
%\hline  
9 & 45 & 23 & 44 & 32 & 45 & 30 & 36 & 73\\
%\hline  
10 & 50 & 19 & 48 & 36 & 50 & 39 & 51 & 65\\
%\hline  
11 & 55 & 22 & 52 & 41 & 55 & 47 & 56 & 92\\
%\hline  
12 & 60 & 26 & 59 & 59 & 60 & 51 & 61 & 90\\
%\hline  
13 & 65 & 30 & 64 & 68 & 65 & 63 & 63 & 95\\
%\hline  
14 & 70 & 35 & 67 & 45 & 70 & 68 & 66 & 122\\
%\hline  
15 & 75 & 40 & 73 & 98 & 75 & 75 & 77 & 152\\
%\hline  
16 & 80 & 45 & 79 & 60 & 80 & 89 & 81 & 157\\
%\hline  
17 & 85 & 52 & 84 & 70 & 85 & 98 & 83 & 95\\
%\hline  
18 & 90 & 56 & 87 & 86 & 90 & 110 & 89 & 196\\
%\hline  
19 & 95 & 62 & 92 & 82 & 95 & 128 & 93 & 198\\
%\hline  
20 & 100 & 68 & 97 & 144 & 100 & 138 & 97 & 146\\
\bottomrule
\end{tabular*}
\end{table*}
%%%%%%%%%%%%%%%%%%%%%%%%%%%%%%%%%%%%%%%%%%%%%%%%%%%%%%%%%%%%%